\documentclass[aps,prc,preprint,groupedaddress,showpacs]{revtex4}

\usepackage{graphicx}
\usepackage{amssymb}

\begin{document}


\title{Predictions for two-pion correlations for $\sqrt s=14$ TeV proton-proton collisions}


\author{T. J. Humanic}
\email[]{humanic@mps.ohio-state.edu}
\affiliation{Department of Physics, The Ohio State University,
Columbus, Ohio, USA}


\date{\today}

\begin{abstract}
A simple model based on relativistic geometry and final-state hadronic rescattering
is used to predict pion source parameters extracted in two-pion correlation studies of
proton-proton collisions at $\sqrt s=14$ TeV. By comparing the results of these model studies
with data, it might be possible to obtain information on the hadronization time
in these collisions. As a test of this model, comparisons between existing two-pion
correlation data at $\sqrt{s}= 1.8$ TeV and results from the model are made. It is
found at this lower energy that using a short hadronization time in the model best describes
the trends of the data.
\end{abstract}


\maketitle


\section{Introduction}
With first proton-proton collisions at $\sqrt{s}= 14$ TeV from the Large Hadron Collider (LHC)
being only about a year (or so) away, it is tempting to use simple models to make baseline
predictions of what we might expect for "bread and butter" observables at this unexplored
energy. Comparisons between data and such models could give us a first impression of
the presence of new physics which might cause significant disagreements between them. If significant
disagreements are seen, the simple models might then be used to point in a direction as to the nature
of the new physics.

The "bread and butter" observable studied in the present work is two-pion correlations. From
this observable, information about the space-time geometry of the pion
emissions produced in the proton-proton collisions can be, at least in principle, extracted
using the interferometric technique pioneered by Hanbury Brown and Twiss
(HBT) \cite{hbt1} and first used in particle physics in $p-\bar{p}$
collisions by Goldhaber, Goldhaber, Lee and Pais (GGLP) \cite{gglp}. Many such experimental
studies using this technique have been carried out over the past nearly 50 years
\cite{Lisa:2005a,Humanic:2006a}, the highest energy study being carried out at the Tevatron with $p-\bar{p}$ collisions at $\sqrt{s}= 1.8$ TeV \cite{e735}. The strategy of the present study is to use a simple model
based on relativistic geometry and final-state hadronic rescattering to predict
pion source parameters extracted in two-pion correlation studies of
proton-proton collisions at $\sqrt s=14$ TeV. As a test of this model, comparisons
with existing two-pion correlation data at $\sqrt{s}= 1.8$ TeV are made. A similar study which
served as the inspiration for the present work has been published by
Paic and Skowronski \cite{Paic:2005a}. The main differences between that study and the present
approach are that in the present approach 1) a somewhat simpler
geometric picture of hadronization is used, e.g. no explicit
identification of jets vs. non-jets is made, 2) for simplicity
only 1-dimensional invariant correlation functions are studied, 3) at Tevatron energy Gaussian fits to the model-generated two-pion correlation functions are made to directly compare with experiment whereas at LHC energy the two-pion correlation function is fit to a more general function , and 4) the effects of final-state hadronic rescattering are included since particle multiplicities become relatively large at these higher energies.

The paper is divided into the following sections: Section II gives a description of the model, Section III presents results of the model and discussions for $p-p$ collisions at $\sqrt{s}= 14$ TeV and 
$p-\bar{p}$ at 1.8 TeV and comparisons with Tevatron data, and Section IV gives conclusions and summary.

\section{Description of the model}
The model calculations are carried out in four main steps: A) generate hadrons in $p-p$ and $p-\bar{p}$ collisions from PYTHIA  \cite{pythia6.3}, B) employ a simple space-time geometry picture for hadronization of the
PYTHIA-generated hadrons,  C) calculate the effects of final-state rescattering among the hadrons,
and D) impose Bose-Einstein correlations pairwise on pions, calculate correlation functions,
and fit the correlation functions with Gaussian or more general functions to extract pion source parameters. These steps will now be discussed in more detail.

\subsection{Generation of the $p-p$ collisions with PYTHIA}
The $p-p$ and $p-\bar{p}$ collisions were modeled with the PYTHIA code \cite{pythia6.3}, version 6.326. The parton distribution functions used were the same as used in Ref. \cite{bh}. Events were generated
in ``minimum bias'' mode, i.e. setting the low-$p_T$ cutoff for parton-parton collisions to zero (or
in terms of the actual PYTHIA parameter, $ckin(3)=0$). Runs were made both with $\sqrt{s}= $ 1.8
and 14 TeV to simulate Tevatron and LHC (full energy) collisions, respectively. Information saved
from a PYTHIA run for use in the next step of the procedure were the momenta and identities
of the ``direct'' (i.e. redundancies removed) hadrons (all charge states) $\pi$, $K$, $p$, $n$,
$\Lambda$, $\rho$, $\omega$, $\eta$, ${\eta}'$, $\phi$, and $K^*$. These particles were
chosen since they are the most common hadrons produced and thus should have the biggest
effect on the two-pion correlation functions extracted in these calculations.

\subsection{The space-time geometry picture for hadronization}
The simple space-time geometry picture for hadronization consists of the emission of a PYTHIA
particle from a thin uniform disk of radius 1 fm in the plane transverse (x-y) to the beam direction (z) followed by
its hadronization which occurs in the proper time of the particle, $\tau$. The space-time
coordinates at hadronization in the lab frame $(x_h, y_h, z_h, t_h)$ for a particle with momentum
coordinates $(p_x, p_y, p_z)$, energy $E$, rest mass $m_0$, and transverse disk
coordinates $(x_0, y_0)$ can then be written as

\begin{eqnarray}
x_h = x_0 + \tau \frac{p_x}{m_0} \\
y_h = y_0 + \tau \frac{p_y}{m_0} \\
z_h = \tau \frac{p_z}{m_0} \\
t_h = \tau \frac{E}{m_0}
\end{eqnarray}

The simplicity of this geometric picture is now clear: it is just an expression of causality with the
assumption that all particles hadronize with the same proper time, $\tau$. A similar hadronization
picture (with an initial point source) has been applied to $e^+-e^-$ collisions\cite{csorgo}.
We do not
a priori know the value of $\tau$ but from the geometric
scale of a $p-p$ collision we might guess that $\tau$ falls in the
range $0<\tau<\sim 1$ fm/c. In order to study the dependence of the results of the
model on this parameter, calculations will be carried out with a range of values.
Note that the HBT results given later from the model are found to not strongly depend on the
choice of the radius of the initial transverse disk within a range of $1\pm0.5$ fm or
on the choice of a disk versus a smoothly dropping off distribution such as a Gaussian due
to the effects of these assumptions being ``washed out'' by the randomizing effects of the
``causality term'' in Eqs.(1) and (2) and of final-state rescattering.

\subsection{Final-state hadronic rescattering}
Since very high energy $p-p$ collisions are being considered here and the
most interesting collisions are normally those producing the highest
particle multiplicities, it seems possible that at early times during the collision the
particle density could reach a level at which significant final-state hadronic rescattering
might take place. An attempt is made to take this effect into account in the present calculations.

The hadronic rescattering calculational method used is similar to that
employed in previous calculations for heavy-ion collisions at CERN Super Proton
Synchrotron (SPS) energies  and BNL Relativistic Heavy Ion Collider (RHIC)
energies \cite{Humanic:1998a,Humanic:2006a}, where particle densities are high enough to produce
significant rescattering effects. Rescattering is simulated with a semi-classical Monte Carlo
calculation which assumes strong binary collisions between hadrons.
Relativistic kinematics is used throughout. The hadrons input into the
calculation from PYTHIA are pions, kaons,
nucleons and lambdas ($\pi$, K,
N, and $\Lambda$), and the $\rho$, $\omega$, $\eta$, ${\eta}'$,
$\phi$, $\Delta$, and $K^*$ resonances. For simplicity, the
calculation is isospin averaged (e.g. no distinction is made among a
$\pi^{+}$, $\pi^0$, and $\pi^{-}$).

The rescattering calculation finishes
with the freeze out and decay of all particles. Starting from the
initial stage ($t=0$ fm/c), the positions of all particles in each event are
allowed to evolve in time in small time steps ($\Delta t=0.1$ fm/c)
according to their initial momenta. At each time step each particle
is checked to see a) if it has hadronized ($t>t_h$, where $t_h$ is given in
Equation 4.), b) if it
decays, and c) if it is sufficiently close to another particle to
scatter with it. Isospin-averaged s-wave and p-wave cross sections
for meson scattering are obtained from Prakash et al.\cite{Prakash:1993a}
and other cross sections are estimated from fits to hadron scattering data
in the Review of Particle Physics\cite{pdg}. Both elastic and inelastic collisions are
included. The calculation is carried out to 20 fm/c which
allows enough time for the rescattering to finish (as a test, calculations were also carried out to 40 fm/c with no changes in the results). Note that when this cutoff time is reached, all un-decayed resonances are allowed to decay with their natural lifetimes and their projected decay positions and times are recorded. The
rescattering calculation is described in more detail elsewhere
\cite{Humanic:2006a,Humanic:1998a}. The validity of the numerical
methods used in the rescattering code have recently been studied and
verified\cite{Humanic:2006b}.

\subsection{Correlation function calculation and fitting}
For the two-pion correlation calculations, the two-pion correlation function is formed
and either a Gaussian or more general function is fitted to it to extract the final fit
parameters.  In the present calculation boson statistics are introduced after
rescattering using a method of pair-wise symmetrization of bosons in
 a plane-wave approximation \cite{Humanic:1986a}.
The final step in the calculation is extracting fit parameters by
fitting a parameterization to the Monte-
Carlo-produced two-pion invariant correlation function, $C(Q_{inv})$, where $Q_{inv}$
is the invariant momentum difference defined as the magnitude of the difference
between the four-momenta of the two pions, i.e. $Q_{inv}=\mid p_1-p_2\mid$.
The forms of the Gaussian and general fit functions are given, respectively, by

\begin{eqnarray}
C(Q_{inv}) = A [1+\lambda \exp(- R^2 Q_{inv}^2)]
\end{eqnarray}
or,
\begin{eqnarray}
C(Q_{inv}) = A [1+\lambda \cos{(\alpha Q_{inv}^2)} \exp(- R^\delta Q_{inv}^\delta)]
\end{eqnarray}

where $R$ is a radius parameter, $\lambda$ is an empirical parameter
normally employed to help fit the function to the correlation function
(i.e. $\lambda=1$ in the ideal case of pure Bose-Einstein correlations), $\alpha$ describes oscillations in the correlation function, $\delta$ represents the degree to which the correlation function falls off with increasing $Q_{inv}$, and $A$
is a normalization factor. For the Gaussian case, a simple connection can be made between $R$ and the space-time
distribution of the pion source, $\rho(r)$, where $r$ is a position variable, via
\begin{eqnarray}
\rho(r) \sim \exp(-\frac{r^2}{2 R^2})
\end{eqnarray}
and,
\begin{eqnarray}
C(Q_{inv}) \sim 1+\lambda \mid \tilde{\rho}(Q_{inv})\mid ^2
\end{eqnarray}
where $\tilde{\rho}(Q_{inv})$ is the Fourier Transform of $\rho(r)$ in terms
of $Q_{inv}$. Inserting the Fourier Transform
of Eq.(7) into Eq.(8) gives Eq.(5).
The Gaussian function was used by E735 to extract $R$ and $\lambda$ from
data and thus is used exclusively to extract these parameters in the present calculations
for $p-\bar{p}$ at $\sqrt{s}=1.8$ TeV
to compare with the data.
The general fit function, Eq.(6), is used exclusively to extract fit parameters to the model correlation functions for calculations of $\sqrt{s}=14$ TeV $p-p$ collisions. 
This is done since there are no restrictions on the fit function which can be used, the goal being to characterize the correlation function with as good a fit as possible to study the dependencies of the fit parameters on various kinematical conditions. The motivation for the particular form 
of Eq. (6) is discussed in detail elsewhere\cite{csorgo}. Note also that a form similar to
this has been used to fit preliminary pion correlation functions obtained in the LEP L3 $e^+ - e^-$
experiment in which a hint of a baseline oscillation has been observed \cite{l3}.

\section{Results and Discussion}
Results from the model calculations described above are now presented and discussed. A comparison
between model calculations for $p-\bar{p}$ collisions at $\sqrt{s}=1.8$ TeV and experimental results
from the Tevatron E735 experiment are presented first as a reality check on the present model near the highest energy
collisions presently available, followed by predictions
from the model for $p-p$ collisions at $\sqrt{s}=14$ TeV.

\subsection{Comparisons with data for $p-\bar{p}$ collisions at $\sqrt{s}= 1.8$ TeV}
Although experimental two-pion HBT results for $p-p$ collisions at $\sqrt{s}=14$ TeV
are not yet available from the LHC with which to compare the predictions
which will be presented later in this work,
it is possible to compare results calculated from the present model with existing
experimental two-pion HBT results from Tevatron experiment E735\cite{e735}, which
studied $p-\bar{p}$ collisions at $\sqrt{s}=1.8$ TeV. Such a comparison with $p-\bar{p}$
collision data near the highest existing energy will point towards what expectations
we should have for the present simple model to predict the higher LHC full-energy HBT
behavior.

To carry out this comparison, calculations were made with the present model for $\sqrt{s}=1.8$
TeV $p-\bar{p}$ collisions using the same parton distribution functions in PYTHIA as for the
$\sqrt{s}=14$ TeV case as mentioned above (this was done to be as consistent as possible with
the $\sqrt{s}=14$ TeV calculations -- it is not expected that using
different pdf's in the model calculations would effect the present results significantly) .
Gaussian fit parameters were extracted from the
calculations using Eq.(5) since this was essentially the same fitting procedure used by E735 to
extract the fit parameters $R$ and $\lambda$. The E735 parameters with which comparison
is made in the present work were obtained directly from Table II of Ref.\cite{e735} for the $N_C$
(see below) dependence and from Table III in the same reference (using their
conversion $R=0.254+1.023R_G$, where $R_G$ is defined in the reference) 
for the $p_T$ dependence.
The E735 pion acceptance was simulated in
the model calculations with simple kinematical cuts on rapidity and $p_T$.  Dependency on
the charged particle multiplicity in the E735 hodoscope, $N_C$, was also studied, being
simulated in the model calculations with an acceptance cut\cite{e735}.

\begin{figure}
\begin{center}
\includegraphics[width=140mm]{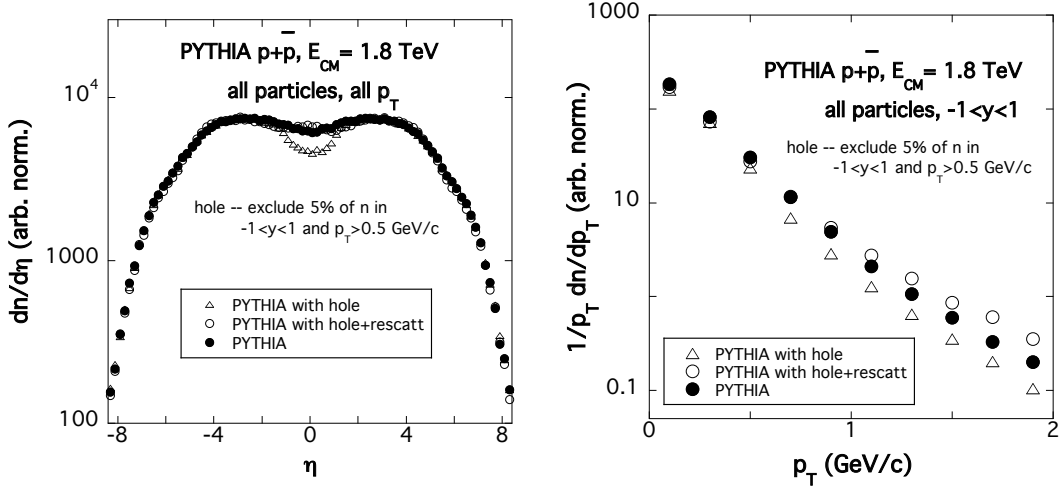} \caption{Pseudorapidity and $p_T$ distributions
from PYTHIA for $p-\bar{p}$ collisions at $\sqrt{s}=1.8$ TeV for three cases: 
1) directly from PYTHIA, 2) PYTHIA with $\eta-p_T$ ``hole'', and 3) same as 2) but with
rescattering turned on and $\tau=0.1$ fm/c. }
\label{fig1}
\end{center}
\end{figure}

\begin{figure}
\begin{center}
\includegraphics[width=140mm]{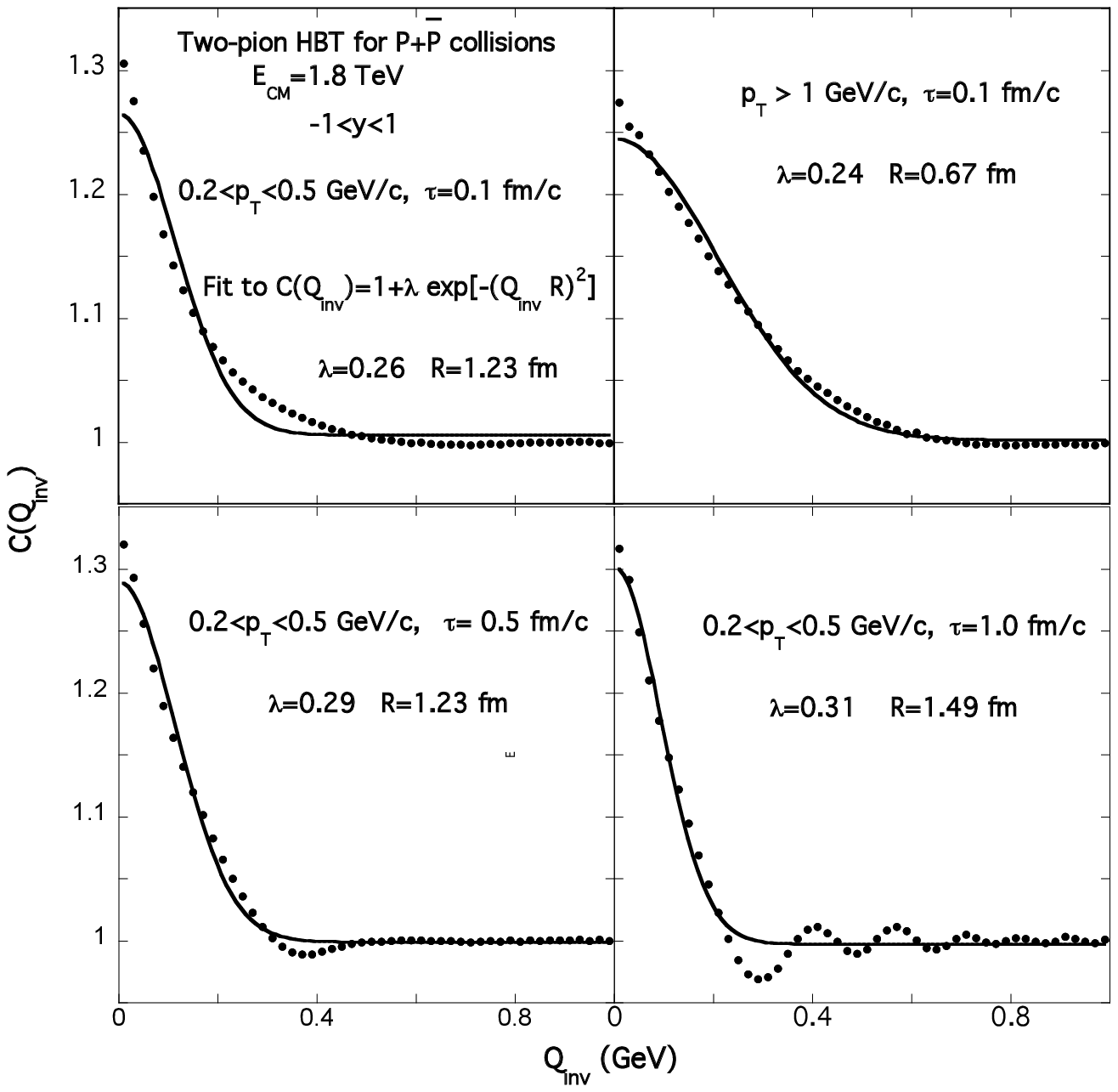} \caption{Sample two-pion correlation
functions from the rescattering model for $\tau=$ 0.1, 0.5, and 1.0 fm/c with fits to the
Gaussian function for $\sqrt{s}=1.8$ TeV $p-\bar{p}$ collisions. }
\label{fig2}
\end{center}
\end{figure}

\begin{figure}
\begin{center}
\includegraphics[width=140mm]{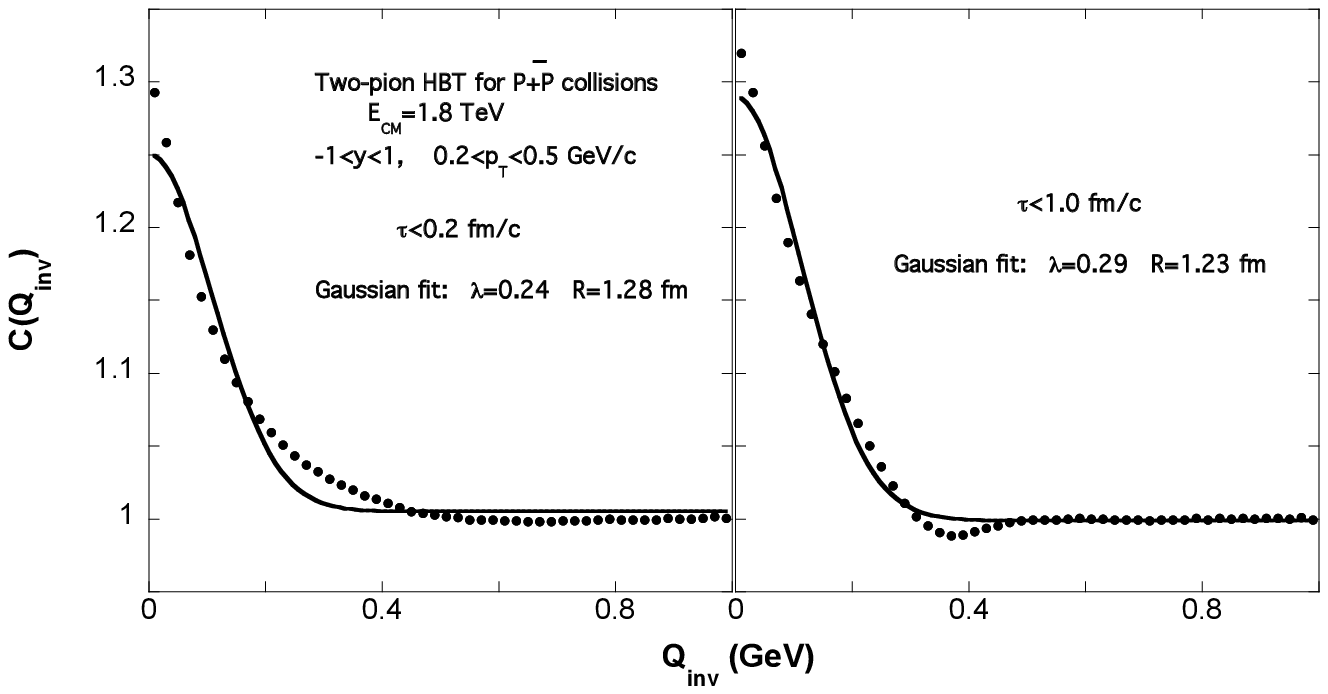} \caption{Sample two-pion correlation
functions from the rescattering model for $\tau<0.2$ and $\tau<1.0$ fm/c with fits to the
Gaussian function for $\sqrt{s}=1.8$ TeV $p-\bar{p}$ collisions. }
\label{fig3}
\end{center}
\end{figure}

\begin{figure}
\begin{center}
\includegraphics[width=140mm]{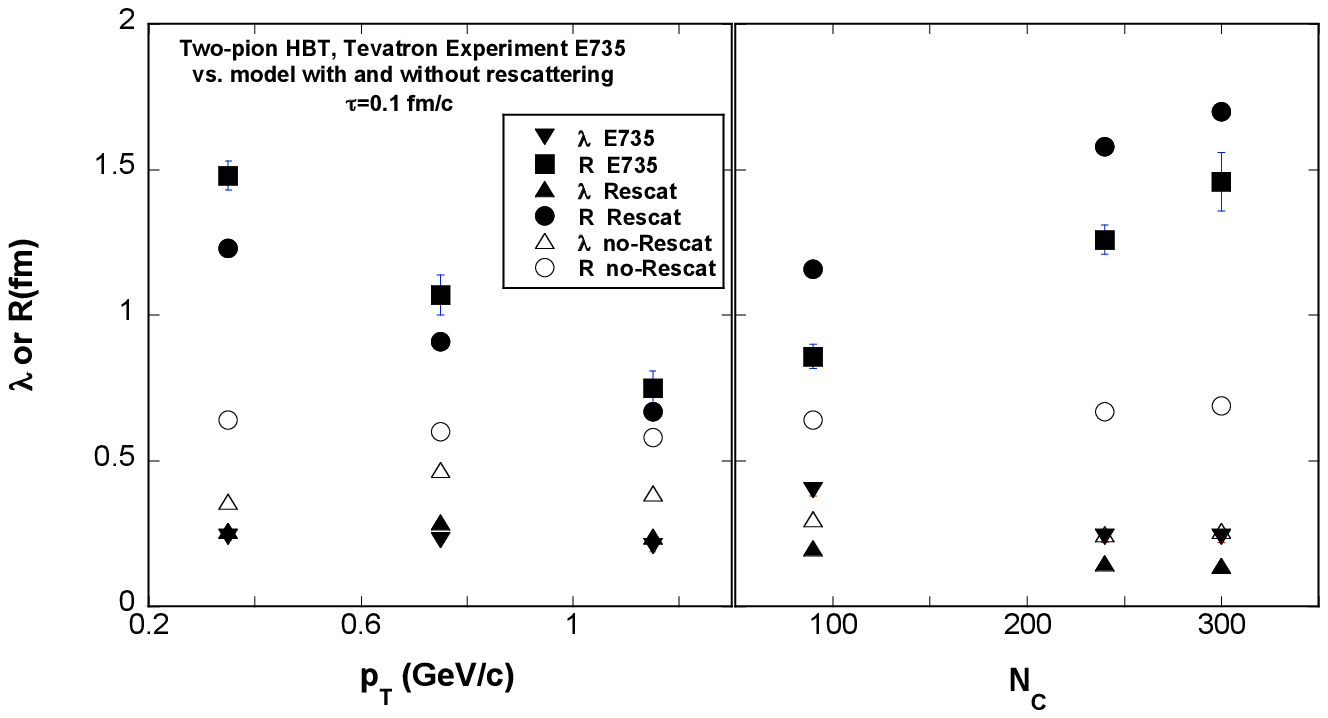} \caption{Comparisons of
Gaussian fit parameters for pions between Tevatron data (Experiment
E735) and model predictions with and without rescattering at $\tau=0.1$ fm/c 
versus $p_T$ and $N_C$.} \label{fig4}
\end{center}
\end{figure}

\begin{figure}
\begin{center}
\includegraphics[width=140mm]{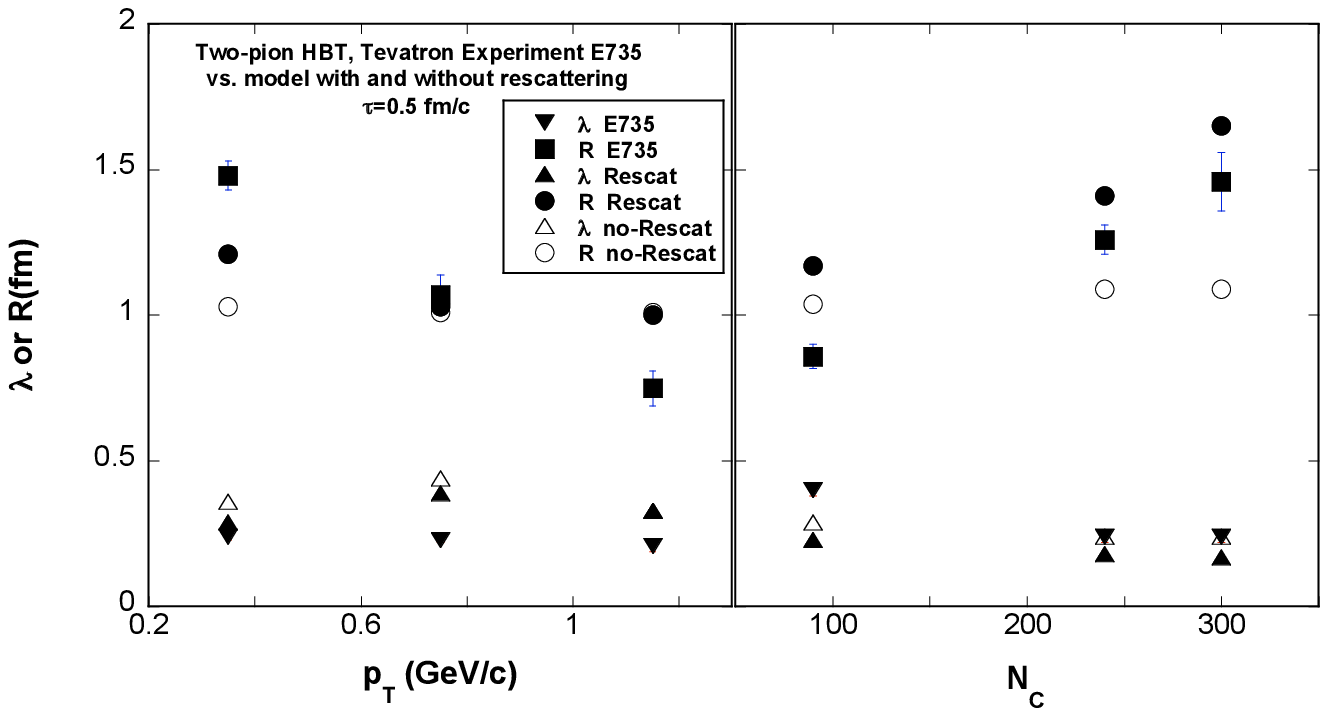} \caption{Comparisons of
Gaussian fit parameters for pions between Tevatron data (Experiment
E735) and model predictions with and without rescattering at $\tau=0.5$ fm/c 
versus $p_T$ and $N_C$.} \label{fig5}
\end{center}
\end{figure}

\begin{figure}
\begin{center}
\includegraphics[width=140mm]{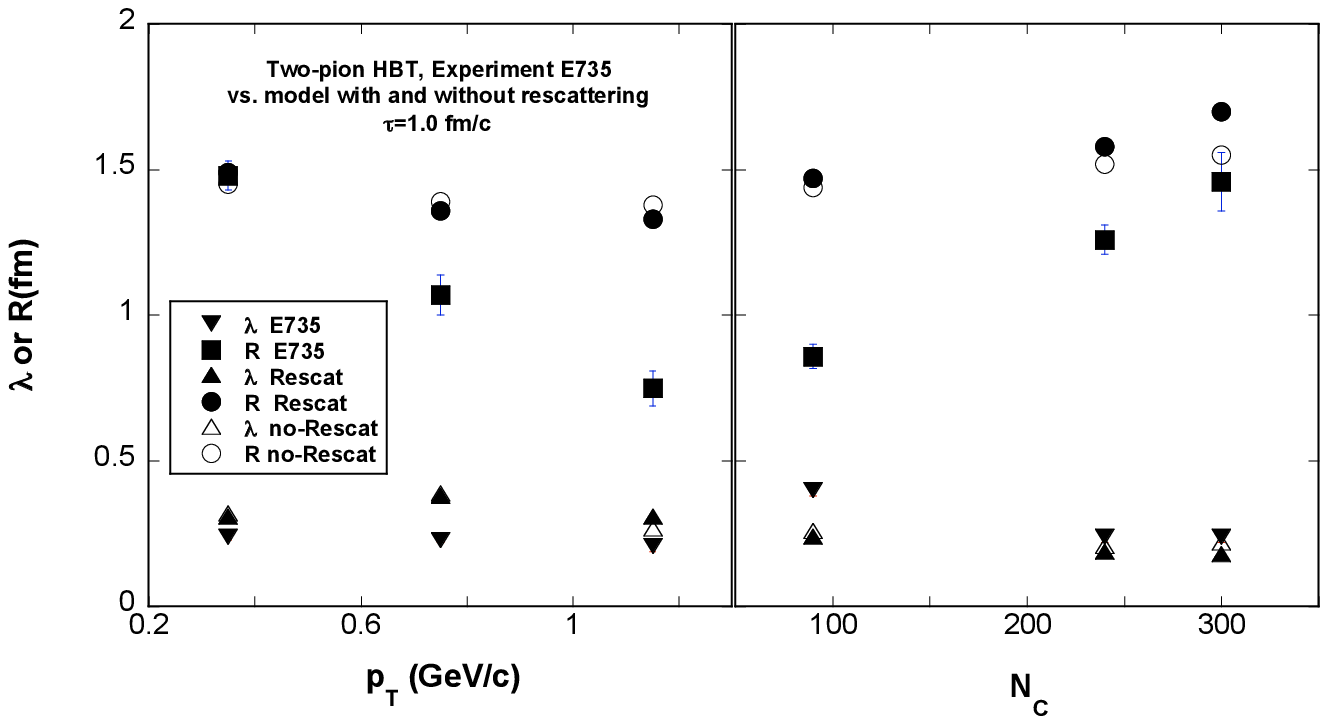} \caption{Comparisons of
Gaussian fit parameters for pions between Tevatron data (Experiment
E735) and model predictions with and without rescattering at $\tau=1.0$ fm/c 
versus $p_T$ and $N_C$.} \label{fig6}
\end{center}
\end{figure}

\begin{figure}
\begin{center}
\includegraphics[width=140mm]{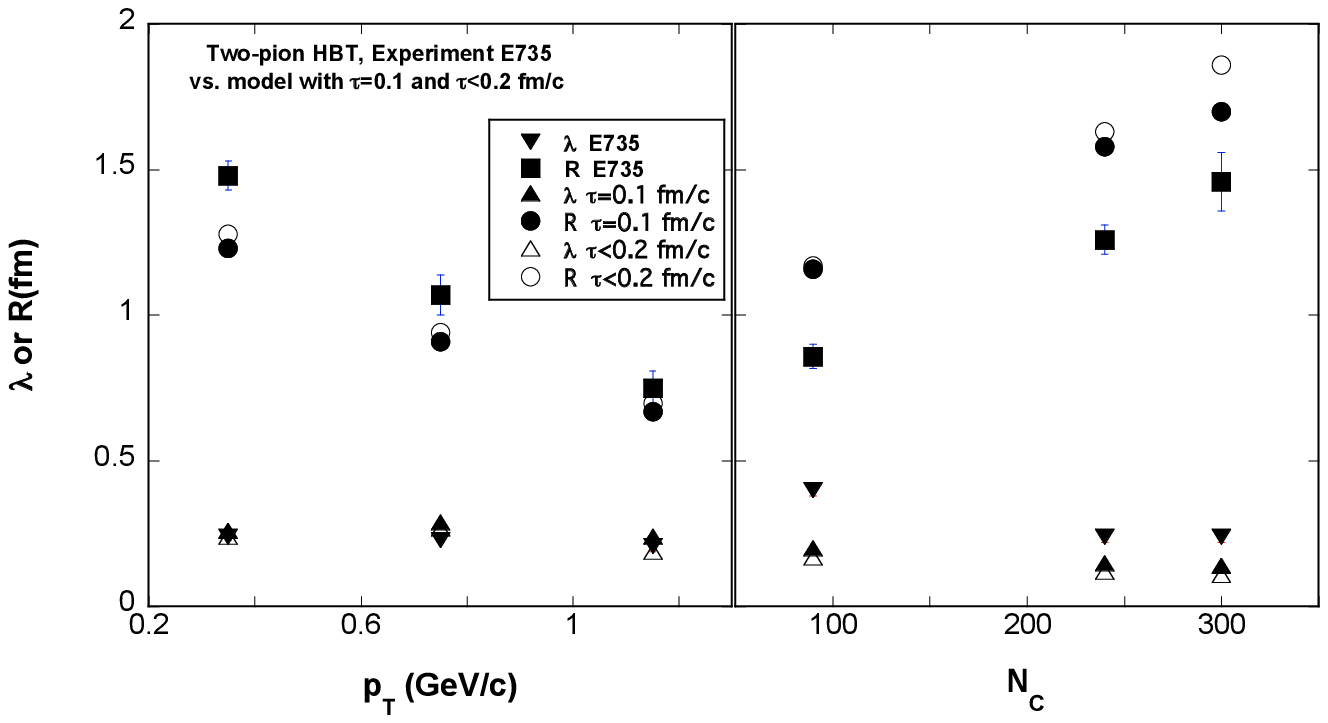} \caption{Comparisons of
Gaussian fit parameters for pions between Tevatron data (Experiment
E735) and model predictions with rescattering for $\tau=0.1$ fm/c and
$\tau<0.2$ fm/c versus $p_T$ and $N_C$.} \label{fig7}
\end{center}
\end{figure}

\begin{figure}
\begin{center}
\includegraphics[width=140mm]{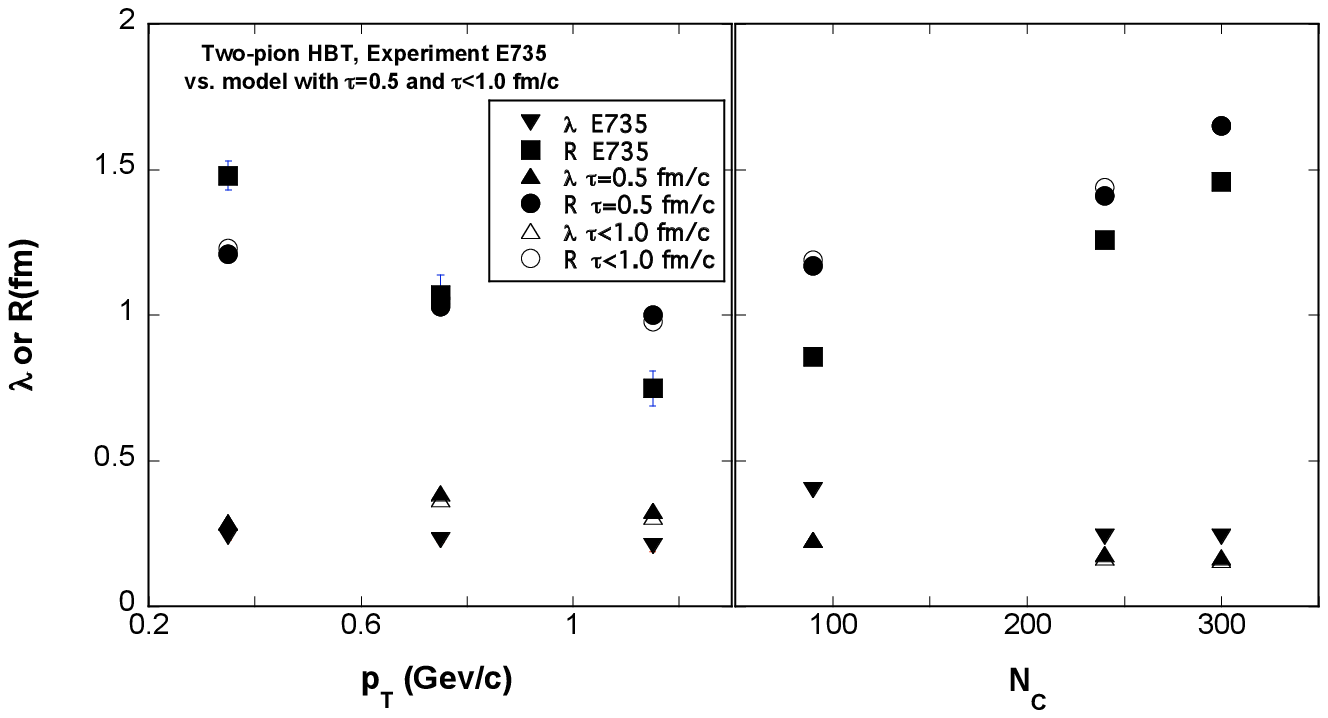} \caption{Comparisons of
Gaussian fit parameters for pions between Tevatron data (Experiment
E735) and model predictions with rescattering for $\tau=0.5$ fm/c and 
$\tau<1.0$ fm/c versus $p_T$ and $N_C$.} \label{fig8}
\end{center}
\end{figure}

Figures \ref{fig1} - \ref{fig8} present results of the model for $\sqrt{s}=1.8$
TeV $p-\bar{p}$ collisions. Comparisons with E735 Gaussian fit parameters are shown in
Figures \ref{fig4} - \ref{fig8}.

Figure \ref{fig1} shows model rapidity and $p_T$ distribution plots for all final-state 
particles (i.e. pions, kaons, and nucleons) for three cases: 1) directly from PYTHIA, 
2) PYTHIA with a $\eta-p_T$ ``hole'', and 3) same as 2) but with rescattering turned on
and $\tau=0.1$ fm/c.
For case 1), PYTHIA events are directly run through the model code without any other
process applied to them except to decay $\Lambda$s, and the $\rho$, $\omega$, $\eta$, ${\eta}'$,
$\phi$, $\Delta$, and $K^*$ resonances, i.e. ``pure PYTHIA''. Since PYTHIA has been tuned
to agree reasonably well with existing experimental data, including Tevatron data, these
distributions should remain at least approximately the same after rescattering has been turned on.
This turns out not to be the case for $\tau= 0.1$ fm/c, since It is found that if ``pure'' PYTHIA events are input into the calculation with rescattering turned on,
a small peak results in the $\eta$ distribution near midrapidity  and the $p_T$ distribution is
overly enhanced compared with ``pure'' PYTHIA. This gives the first suggestion that final-state
hadronic rescattering can play a noticeable role in these collisions. In an effort to compensate
for the rescattering effects so as to give approximate agreement with the ``pure'' PYTHIA
distributions, a $\eta-p_T$ ``hole'' is inserted in the input PYTHIA events before rescattering,
as shown, and the rescattering then fills the ``hole'' to approximately agree with the ``pure'' PYTHIA
distributions, also shown in Figure \ref{fig1}. For this case, the ``hole'' is defined by randomly
throwing away 5\% of the particles in the input PYTHIA events in the $y-p_T$ region $-1<y<1$ and $p_T>0.5$ GeV/c. For the larger values of $\tau$ studied, i.e. $\tau=0.5$ and $1.0$ fm/c, less
rescattering takes place due to the larger initial hadronization volume and thus lower initial
particle density and the ``hole'' depth is reduced, using the prescription $5\%(0.1/\tau)$
to define it. The justification for using this ``hole'' method is that reasonable agreement
with the ``pure'' PYTHIA distributions is obtained with recattering turned on. Note that including
or not including the ``hole'' has only a small effect on the HBT results presented later.

Figure \ref{fig2} shows sample two-pion correlation
functions from the rescattering model for $\tau=$ 0.1, 0.5, and 1.0 fm/c with fits to the
Gaussian function, Eq. (5),  for $\sqrt{s}=1.8$ TeV $p-\bar{p}$ collisions. A comparison is also
made for the $\tau=0.1$ fm/c case between two $p_T$ cuts on the pions, i.e.
$0.2<p_T<0.5$ GeV/c and $p_T>1$ GeV/c. As seen, the Gaussian fits qualitatively reproduce
the trends of the model correlation functions, but do not well represent all of the details of
the shapes, which include an exponential-like shape for $\tau=0.1$ fm/c and some oscillatory
behavior for $\tau=0.5$ and $1.0$ fm/c. The oscillatory behavior is a feature of the delta-function
assumption of $\tau$, which becomes more prominent for larger values of $\tau$ \cite{csorgo}.

Figure \ref{fig3} shows sample correlation functions where the model is run with a uniform
distribution of $\tau$ as a test, for the two cases $\tau<0.2$ fm/c and $\tau<1.0$ fm/c. 
As seen by comparing Figures \ref{fig2} and \ref{fig3}, these cases closely resemble the correlation functions for $\tau=0.1$ and $\tau=0.5$ fm/c, respectively. A more complete comparison
is shown later.

Figure \ref{fig4}-\ref{fig6} show comparisons of
Gaussian fit parameters for pions between Tevatron data (Experiment
E735) and model predictions with and without rescattering at $\tau=0.1, 0.5$ and $1.0$ fm/c,
respectively, versus $p_T$ and $N_C$. As anticipated earlier, rescattering is seen to have
the greatest effect on the fit parameters for the smallest value of $\tau$, $\tau=0.1$ fm/c,
becoming less important as $\tau$ increases until it is seen to have an almost
negligible effect at $\tau=1.0$ fm/c. All three $\tau$ cases (with rescattering) 
do an adequate job of describing the flat dependence of $\lambda$ on 
$p_T$ and $N_C$ seen in E735. It is also seen that the overall trends of the
E735 $R$ dependencies, i.e. decreasing with increasing $p_T$ and
increasing with increasing $N_C$, are best reproduced by the $\tau=0.1$ fm/c case
with rescattering turned on, the larger $\tau$ model predictions becoming 
progressively flatter with increasing $\tau$. Another feature found in the
correlation functions of the higher $\tau$ values not found in the E735 correlation
functions is the oscillation in the baseline seen in Figure \ref{fig2}. No such oscillation
appears for the $\tau=0.1$ fm/c case, in agreement with E735.

Figures \ref{fig7} and \ref{fig8} show results for running the model with flat distributions
of $\tau$, i.e. $\tau<0.2$ fm/c and $\tau<1.0$ fm/c, and comparing with the delta-function
cases $\tau=0.1$ and $0.5$ fm/c, which are the average values of the two flat ranges,
respectively, with rescattering turned on, and compared with E735. As seen, the fit
parameters for the flat $\tau$ distributions give virtually the same results as the 
delta-function $\tau$ distributions, demonstrating that either method of running the
model gives almost identical results.

Summarizing this section, the main result of the comparison of the HBT fit 
parameters from the present 
model with those from Tevatron experiment E735 shown above is that the $\tau=0.1$ fm/c case
best describes the trends of the fit parameters on $p_T$ and $N_C$. This would seem
to imply that the hadronization time in these collisions is short, i.e. $\tau << 1$ fm/c,
and, as a consequence of this, large hadron densities exist at the early stage of the
collision resulting in significatnt hadronic rescattering effects.

\subsection{Predictions for p-p collisions at $\sqrt{s}= 14$ TeV}
\begin{figure}
\begin{center}
\includegraphics[width=140mm]{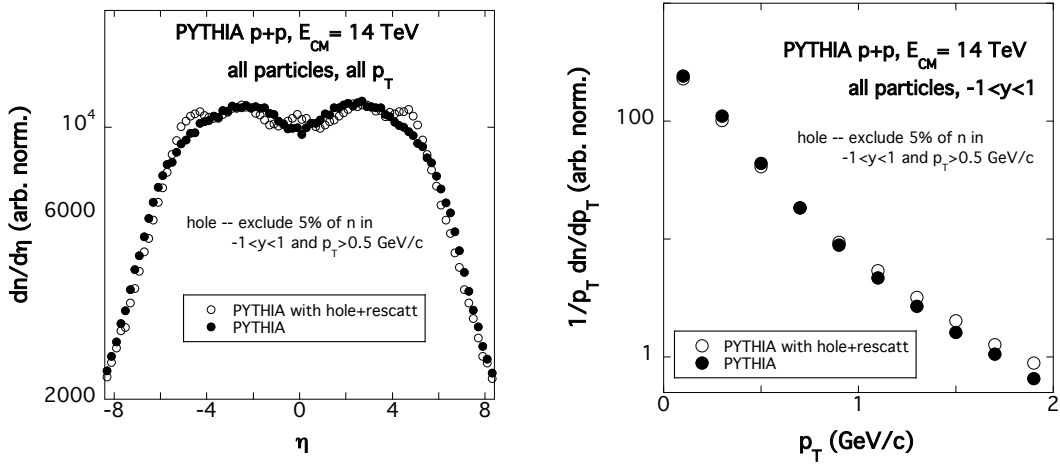} \caption{Pseudorapidity and $p_T$ distributions
from PYTHIA for $p-p$ collisions at $\sqrt{s}=14$ TeV for two cases: 
1) directly from PYTHIA, and 2) PYTHIA with $\eta-p_T$ ``hole'' and with
rescattering turned on for the case $\tau=0.1$ fm/c. }
\label{fig9}
\end{center}
\end{figure}

\begin{figure}
\begin{center}
\includegraphics[width=140mm]{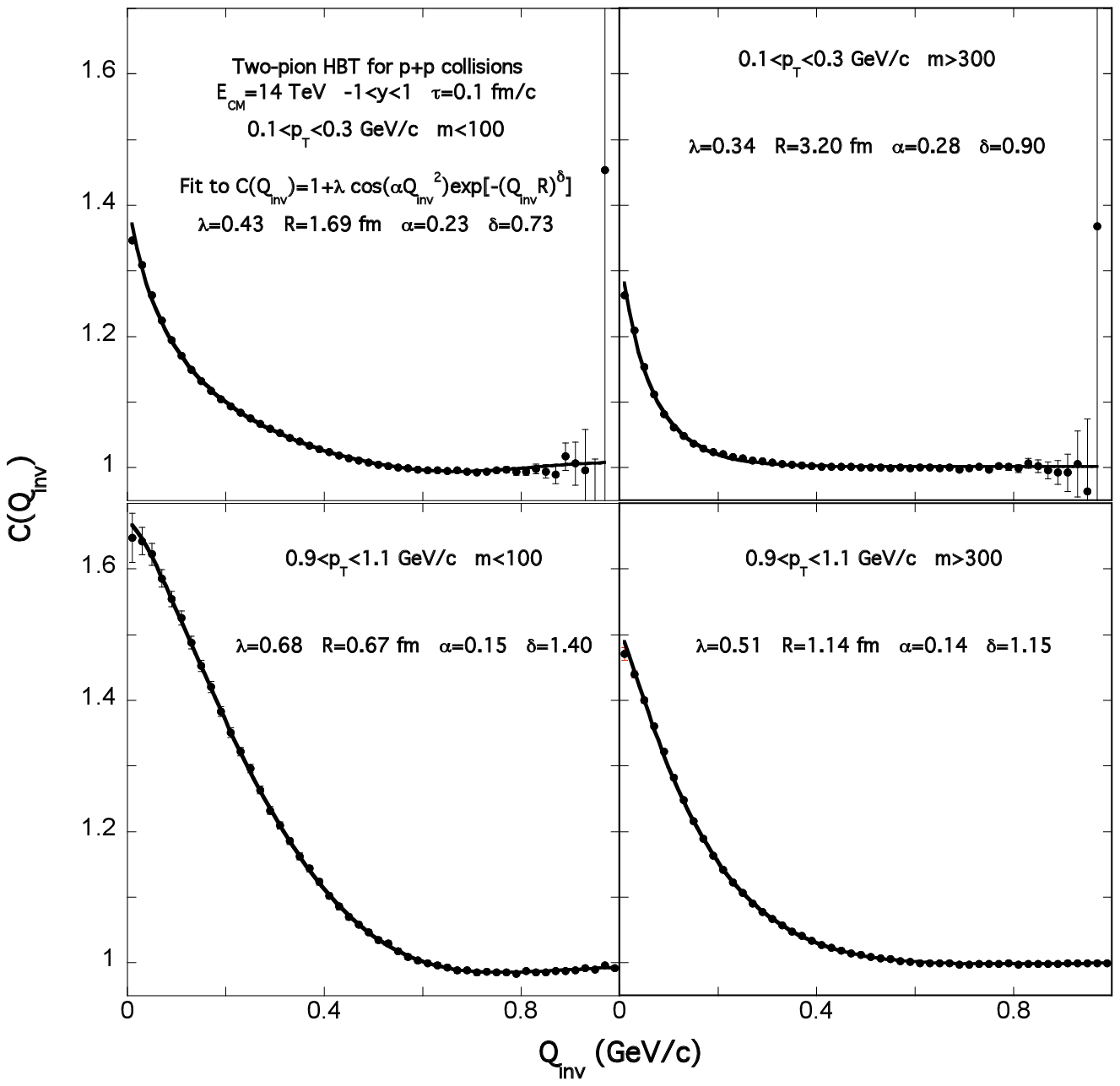} \caption{Sample two-pion correlation
functions from the rescattering model for $\tau=$ 0.1 fm/c with fits to the
general function (i.e. Eq.(6)) for $\sqrt{s}=14$ TeV $p-p$ collisions.}
\label{fig10}
\end{center}
\end{figure}

\begin{figure}
\begin{center}
\includegraphics[width=140mm]{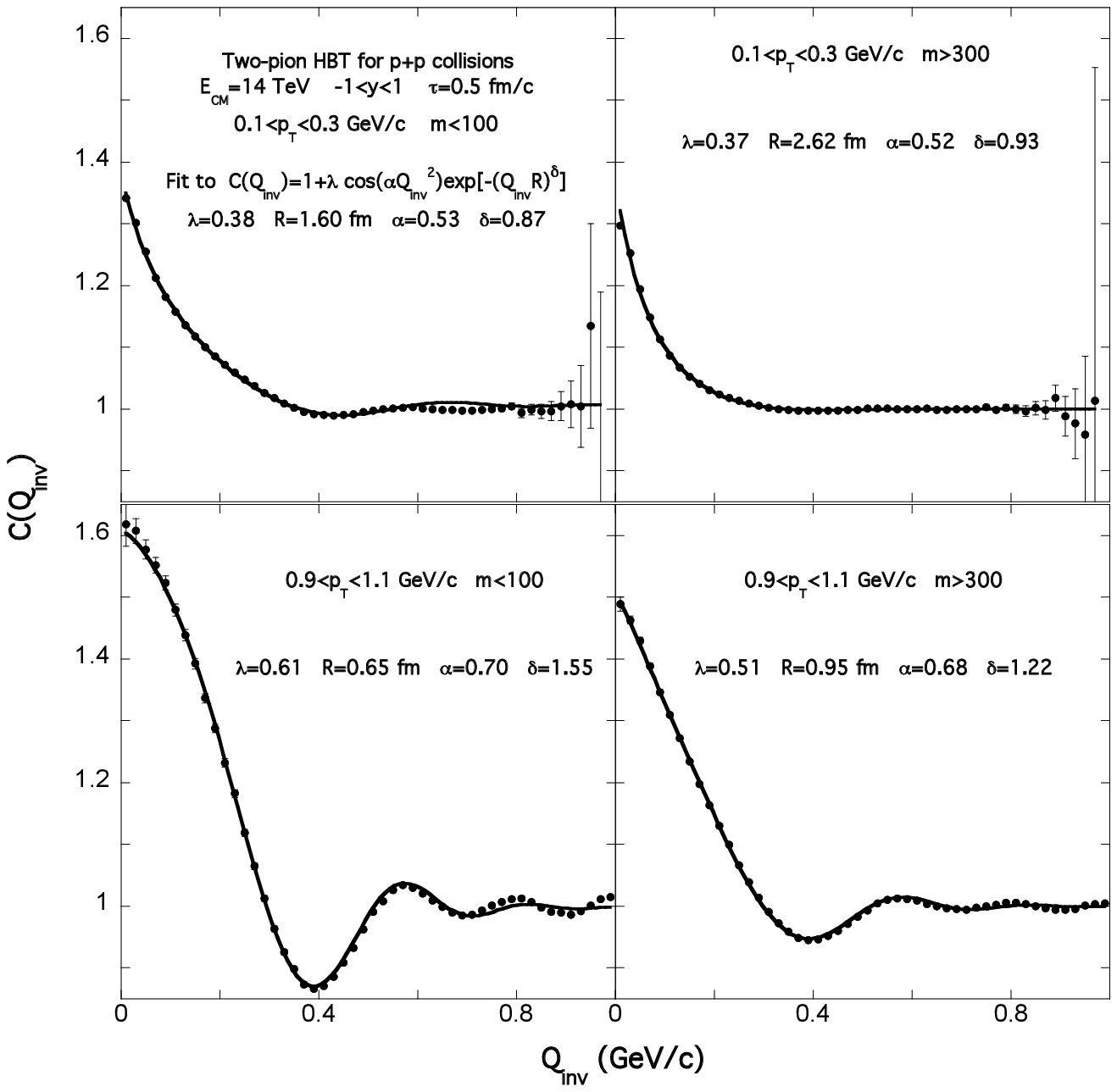} \caption{Sample two-pion correlation
functions from the rescattering model for $\tau=$ 0.5 fm/c with fits to the
general function (i.e. Eq. (6)) for $\sqrt{s}=14$ TeV $p-p$ collisions.}
\label{fig11}
\end{center}
\end{figure}

\begin{figure}
\begin{center}
\includegraphics[width=140mm]{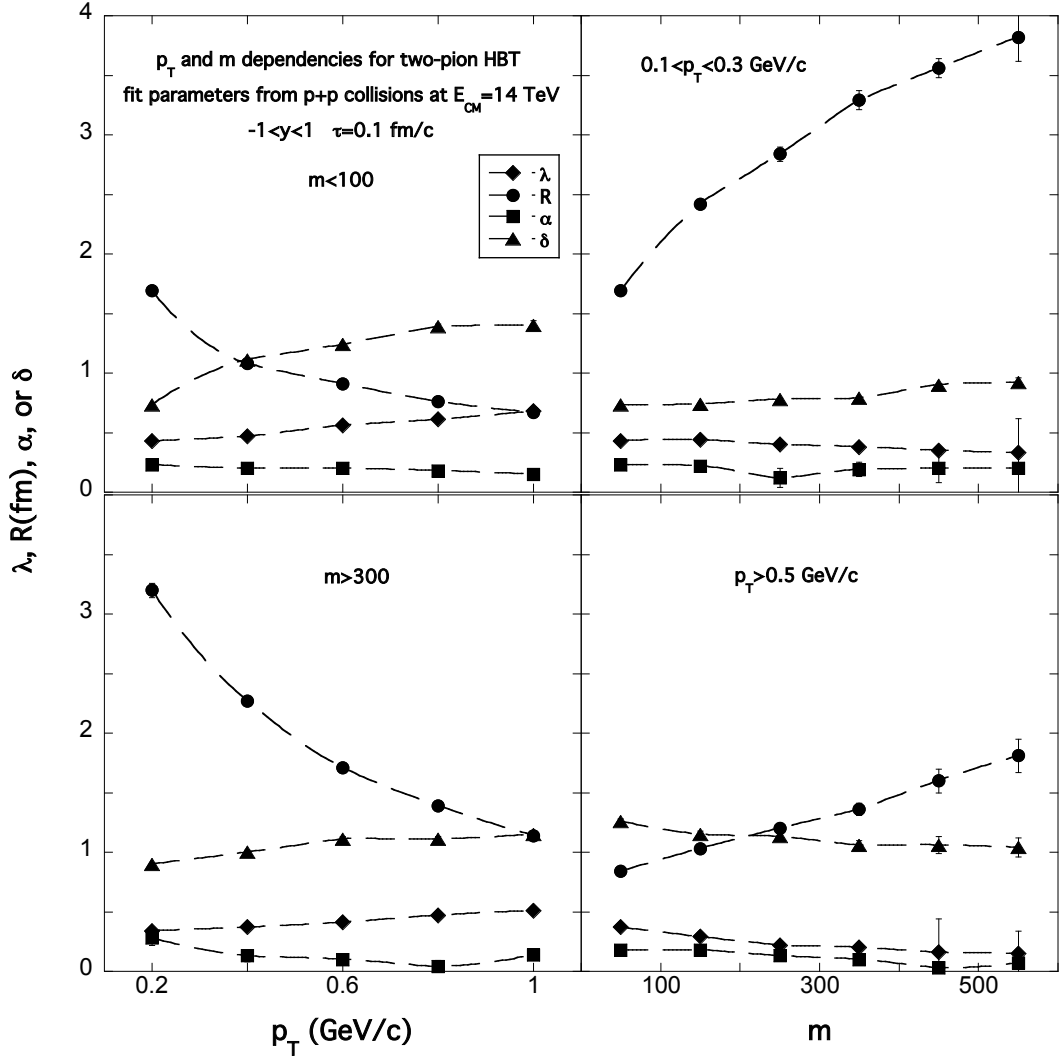} \caption{General function fit parameters
versus $p_T$ and particle multiplicity from the rescattering model with $\tau=0.1$ fm/c
for several multiplicity and $p_T$ cuts for $p-p$ collisions at $\sqrt{s}=14$ TeV. 
The dashed lines are drawn to guide the eye.}
\label{fig12}
\end{center}
\end{figure}

\begin{figure}
\begin{center}
\includegraphics[width=140mm]{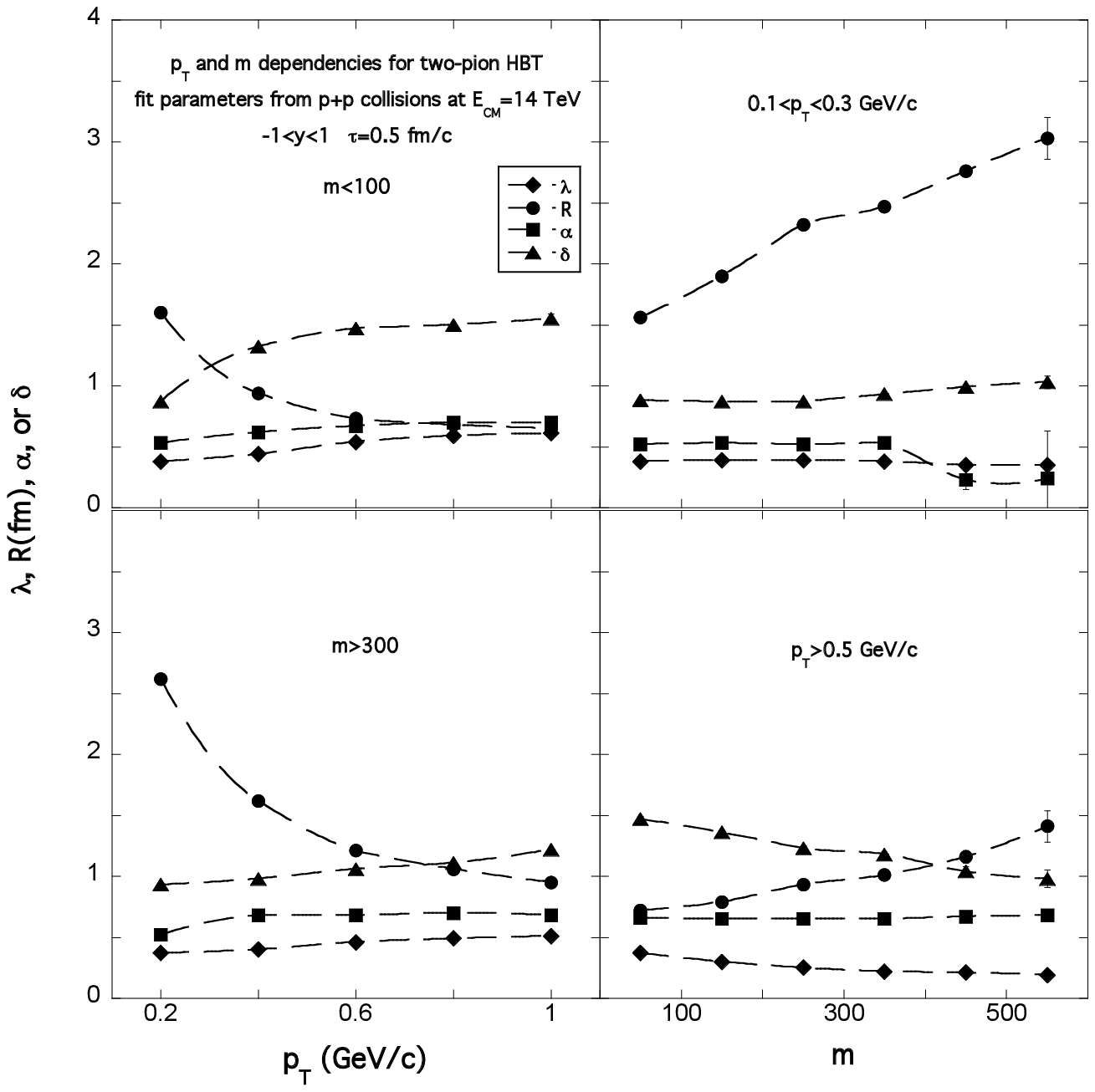} \caption{General function fit parameters
versus $p_T$ and particle multiplicity from the rescattering model with $\tau=0.5$ fm/c
for several multiplicity and $p_T$ cuts for $p-p$ collisions at $\sqrt{s}=14$ TeV. 
The dashed lines are drawn to guide the eye.}
\label{fig13}
\end{center}
\end{figure}

Figures \ref{fig9} - \ref{fig13} present results from the present model calculations for
$p-p$ collisions at $\sqrt{s}=14$ TeV to simulate full-energy LHC collisions. All calculations
shown are for pions at mid-rapidity, i.e. $-1<y<1$. Several values of $\tau$ and cuts on
particle multiplicity and pion $p_T$ are studied. In the results shown below, multiplicity is
defined as the total multiplicity of pions, kaons, and nucleons of all charge states for all
rapidity and $p_T$ in a collision. In order to compare with future experiments, the
approximate correspondence between the total multiplicity bins used and the more
experiment-friendly average detectable ($p_T>100$ MeV/c) mid-rapidity ($-1<y<1$)
charged particle multiplicity is shown in Table I. Also shown in Table I is the fraction
of minimum bias events corresponding to each multiplicity bin. From this it is
seen that all multiplicities used are predicted to be easily experimentally accessible.

\begin{table}
\begin{center}
\caption{Approximate correspondence between the total multiplicity
bins used in the present calculations and the average
detectable mid-rapidity charged multiplicity. The fraction
of minimum bias events is also shown.}
\begin{tabular}{ccr} \\ \hline
Total mult.bin           &    Ave.detectable charged   & fraction of \\
           m                      &    particle  mult.at mid-y      & MB events\\ \hline
0-100          & 5  & 0.42 \\
100-200           & 14 & 0.34 \\
200-300           & 26  & 0.14 \\
300-400           & 41  & 0.069 \\
400-500           & 58  & 0.026 \\
500-600           & 79  & 0.0042 \\
$>$ 300           & 47  & 0.093\\ \hline
\end{tabular}
\end{center}
\end{table}

The following choices were made for the conditions used in the model
calculations in generating the LHC predictions:
\begin{itemize}
\item Based on the comparisons presented above between the model and E735 results, 
predictions using the ``delta-function'' model for $\tau$ for the cases
$\tau=0.1$ and $0.5$ fm/c were made. As shown above for the Tevatron calculations, these
cases give almost identical results as for the ``flat-distribution'' model for
$\tau<0.2$ and $1.0$ fm/c. Although the closest agreement between the present
rescattering calculations and E735 results was obtained for the $\tau=0.1$ fm/c case,
predictions are also included for $\tau=0.5$ fm/c since the hadronization time for
$\sqrt{s}=14$ TeV collisions may be larger than that for $\sqrt{s}=1.8$ TeV. 
\item Employ the same ``$\eta-p_T$ hole'' method as for the Tevatron calculations,
i.e. the ``hole'' is defined by randomly
throwing away 5\% of the particles in the input PYTHIA events in the $y-p_T$ region $-1<y<1$ and $p_T>0.5$ GeV/c for the $\tau=0.1$ fm/c case, and scaled down accordingly for the
$\tau=0.5$ fm/c case. Comparisons of pseudorapidity and $p_T$ distributions 
between PYTHIA run for the maximum LHC energy and PYTHIA with the ``hole''
and rescattering turned on for $\tau=0.1$ are shown in Figure \ref{fig9}.
\item To extract the pion HBT fit parameters from the invariant correlation functions generated
by the model, use Eq.(6). As described above, this is to better characterize the finer features
of the correlation functions and thus get better fits to the calculations. Figures \ref{fig10} and
\ref{fig11} show fits to sample model-generated correlation functions for the cases
$\tau=0.1$ and $0.5$ fm/c, respectively. As seen, the fits are in general quite good.
\end{itemize}

Figures \ref{fig12} and \ref{fig13} show the predicted dependences of the fit parameters
on $p_T$ and total multiplicity, $m$, for the cases $\tau=0.1$ and $0.5$ fm/c, respectively
for $\sqrt{s}=14$ TeV $p-p$ collisions. Plots are made with low and high multiplicity cuts, i.e. $m<100$
and $m>300$, and low and high $p_T$ cuts, i.e. $0.1<p_T<0.3$ GeV/c and
$0.9<p_T<1.1$ GeV/c. The behaviors of the fit parameters seen in Figures \ref{fig12}
and \ref{fig13} are discussed separately below.

\subsubsection{$R$-parameter}
The $R$-parameter, which is related to the ``size'' of the pion-emitting source (see Eq.(7)),
is seen to have the largest variation for different kinematical cuts,
i.e. the strongest dependencies on $p_T$ and $m$. These are greatest for $\tau=0.1$ fm/c
since they are almost completely due to rescattering effects. In Figure \ref{fig12}
it is seen that for the proper kinematical cuts $R$ can be made to increase by a factor
of three for increasing $m$ or decrease by a factor of three for increasing $p_T$. Experimental observation of such strong variations in $R$ would be a convincing signature for
the presence of significant rescattering effects and therefore a short hadronization time.

\subsubsection{$\lambda$-parameter}
The $\lambda$-parameter, which is related to the ``strength'' of the HBT effect, is seen to have weak dependences on the kinematical
cuts and to have a similar magnitude for both the $\tau=0.1$ and $0.5$ fm/c cases.
It tends to have a magnitude such that $\lambda<0.5$, which is mostly due to the
presence of long-lived resonances in the model calculations which tend to
suppress it.
Though weak, it has a slightly
increasing tendency for increasing $p_T$ and decreasing tendency for increasing
$m$, which is opposite the directions of the dependencies for $R$. 

\subsubsection{$\alpha$-parameter}
The $\alpha$-parameter, which is seen in Eq.(6) to be related to oscillatory behavior
of the correlation function, is seen to mostly depend on the value of $\tau$. As also
seen in Figures \ref{fig10} and \ref{fig11}, it is small, i.e. $\alpha<0.2$, for $\tau=0.1$ fm/c
and large, i.e. $\alpha \sim 0.5-0.6$, for $\tau=0.5$ fm/c. The connection between
$\alpha$ and $\tau$ for the simple case of a delta-function hadronization time
can indeed be shown to be $\tau \sim \alpha$ \cite{csorgo}. Experimental observation of
oscillations in the correlation function, i.e. large $\alpha$ values, would be evidence
for a larger value of the hadronization time, i.e. $\tau>0.5$ fm/c.

\subsubsection{$\delta$-parameter}
As seen in Eq.(6), the $\delta$-parameter is related to how ``exponential-like'' ($\delta \sim 1$)
or ``Gaussian-like'' ($\delta \sim 2$) the correlation function appears. As with $\lambda$,
it is seen to have weak dependences on the kinematical
cuts, to have a similar magnitude for both the $\tau=0.1$ and $0.5$ fm/c cases, and to
have dependencies which are opposite to the directions of the dependencies for $R$. It tends to
have values in the range $0.7<\delta<1.5$.

\section{Conclusions}
A simple model assuming a uniform hadronization proper time and including final-state
hadronic rescattering has been used to predict two-pion HBT fit parameters for $p-\bar{p}$
collisions at $\sqrt{s}=1.8$ TeV and $p-p$ collisions at 14 TeV. 
For small values of $\tau$, i.e. $\tau<0.5$ fm/c,
it is found that rescattering has a significant influence on the fit parameters. Comparing the model
predictions with $p-\bar{p}$ experimental results at $\sqrt{s}=1.8$ TeV, the closest agreement
is found for small $\tau$, i.e. $\tau\sim 0.1$ fm/c. This suggests that 1) final-state hadronic
rescattering is already important at $\sqrt{s}=1.8$ TeV and 2) hadronization times are
short. 

As is seen in the above figures, there are significant differences in the magnitudes and
dependences on kinematical variables for the general fit parameters evaluated at
different $\tau$-values in the model calculations carried out for $\sqrt{s}=14$ TeV $p-p$
collisions. Comparisons of these results with actual future data from the LHC will be able
to establish a) if this simple model describes the data in even a qualitative way and
b) if so, the scale of the hadronization time in these collisions.

\begin{acknowledgments}
The author wishes to acknowledge financial support from the U.S.
National Science Foundation under grant PHY-0355007, and to acknowledge computing
support from the Ohio Supercomputing Center.
\end{acknowledgments}


\begin{thebibliography}{00}

\bibitem{hbt1}
R. Hanbury Brown and R. Q. Twiss, Nature {\bf 177}, 27 (1956).

\bibitem{gglp}G. Goldhaber, S. Goldhaber, W. Lee, and A. Pais,
Phys. Rev.{\bf 120}, 300 (1960).

\bibitem{Lisa:2005a}
  M.~A.~Lisa, S.~Pratt, R.~Soltz and U.~Wiedemann,
  Ann.\ Rev.\ Nucl.\ Part.\ Sci.\  {\bf 55}, 357 (2005).

\bibitem{Humanic:2006a}
T.~J.~Humanic,
Int.\ J.\ Mod.\ Phys.\ E {\bf 15}, 197 (2006).

\bibitem{e735}
  T.~Alexopoulos {\it et al.},
  Phys.\ Rev.\ D {\bf 48}, 1931 (1993).

\bibitem{Paic:2005a}
  G.~Paic and P.~K.~Skowronski,
  J.\ Phys.\ G {\bf 31}, 1045 (2005).

\bibitem{pythia6.3}
  T.~Sjostrand, L.~Lonnblad, S.~Mrenna and P.~Skands,
  arXiv:hep-ph/0308153.

\bibitem{bh}
  S.~Dimopoulos and G.~Landsberg,
  Phys.\ Rev.\ Lett.\  {\bf 87}, 161602 (2001).

\bibitem{csorgo}
  T.~Csorgo and J.~Zimanyi,
  Nucl.\ Phys.\  A {\bf 512}, 588 (1990).
  
  \bibitem{l3}
  T.~Novak  [L3 Collaboration],
  AIP Conf.\ Proc.\  {\bf 828}, 539 (2006).

\bibitem{Humanic:1998a} T.~J.~Humanic, Phys.\ Rev.\ C {\bf 57}, 866
(1998).

\bibitem{Prakash:1993a}
M.~Prakash, M.~Prakash, R.~Venugopalan and G.~Welke,
Phys. Rept. {\bf 227}, 321 (1993).

\bibitem{pdg}
  W.~M.~Yao {\it et al.}  [Particle Data Group],
  J.\ Phys.\ G {\bf 33}, 1 (2006).

\bibitem{Humanic:2006b}
T.~J.~Humanic, Phys.\ Rev.\ C {\bf 73}, 054902 (2006).

\bibitem{Humanic:1986a}
T. J. Humanic, Phys. Rev. C {\bf 34}, 191 (1986).

\end{thebibliography}
\end{document}